\documentclass{aa}
\usepackage{graphicx}
\usepackage{natbib}
\bibpunct{(}{)}{;}{a}{}{,}   

\newcommand{\beq}{\begin{equation}}
\newcommand{\bea}{\begin{eqnarray}}
\newcommand{\eeq}{\end{equation}}
\newcommand{\eea}{\end{eqnarray}}

\begin{document}
   \title{Turbulence and particle acceleration in collisionless
	supernovae remnant shocks} 

   \subtitle{II- Cosmic-ray transport}

   \author{Alexandre Marcowith\inst{1}
	  \and
          Martin Lemoine\inst{2}
	  \and
	  Guy Pelletier\inst{3}
          }
   \offprints{A. Marcowith}

   \institute{Centre d'\'Etudes Spatiales et du Rayonnement, \\ 
	CNRS, Universit\'e Paul Sabatier Toulouse 3,\\
 	9, avenue du Colonel Roche, F-31028 Toulouse C\'edex, France\\
             \email{Alexandre.Marcowith@cesr.fr}
         \and
	Institut d'Astrophysique de Paris,\\
	UMR 7095 CNRS Universit\'e Pierre \& Marie Curie,\\
	98 bis boulevard Arago,\\
	75014 Paris\\
	\email{lemoine@iap.fr}
	  \and
            Laboratoire d'Astrophysique de Grenoble, \\
	CNRS, Universit\'e Joseph Fourier II, \\
	BP 53, F-38041 Grenoble, France; \\
              and Institut Universitaire de France\\
             \email{Guy.Pelletier@obs.ujf-grenoble.fr}
             }
   \date{Received; accepted }

   \abstract{Supernovae remnant shock waves could be at the origin of
cosmic rays up to energies in excess of the knee ($E\simeq3\cdot
10^{15}\,$eV) if the magnetic field is efficiently amplified by the
streaming of accelerated particles in the shock precursor. This paper
follows up on a previous paper \citep{pell05} which derived the
properties of the MHD turbulence so generated, in particular its
anisotropic character, its amplitude and its spectrum. In the present
paper, we calculate the diffusion coefficients, also accounting for
compression through the shock, and show that the predicted
three-dimensional turbulence spectrum $k_\perp S_{3\rm
d}(k_\parallel,k_\perp)\propto k_\parallel^{-1}k_\perp^{-\alpha}$
(with $k_\parallel$ and $k_\perp$ the wavenumber components along and
perpendicular to the shock normal) generally leads to Bohm diffusion
in the parallel direction.  However, if the anisotropy is constrained
by a relation of the form $k_\parallel \propto k_\perp^{2/3}$, which
arises when the turbulent energy cascade occurs at a constant rate
independent of scale, then the diffusion coefficient loses its Bohm
scaling and scales as in isotropic Kolmogorov turbulence. We show that
these diffusion coefficients allow to account for X-ray observations
of supernova remnants.  This paper also calculates the modification of
the Fermi cycle due to the energy lost by cosmic rays in generating
upstream turbulence and the concomittant steepening of the energy
spectrum. Finally we confirm that cosmic rays can produced an
amplified turbulence in young SNr during their free expansion phase
such that the maximal energy is close to the knee and the spectral 
index is close to $2.3$ in the warm phase of the interstellar medium.

 }

   \maketitle
\section{Introduction}

Supernovae (SN) blastwaves are probable sites of cosmic ray (CR)
acceleration up to energies of a few $10^{14}$ eV as discussed by
\citet{laga83}. The cosmic rays are accelerated through the diffusive
shock acceleration mechanism which invokes upstream and downstream
turbulence and whose efficiency is fully determined by the transport
properties of the cosmic rays in these regions [see reviews by
\citet{drur83}, \citet{jone91}, and more recently by \citet{drur01},
\citet{hill05}].  The maximum cosmic-ray energy attainable in
supernova remnants (SNR) scales as the product of magnetic field and
size of the accelerator; the above Lagage \& C\'esarsky estimate for
the maximum energy assumes that the fluctuating part of the magnetic
field $\delta B$ has saturated at a level of the order of the far
upstream interstellar value $B_\infty$ \citep{mack82}.  However,
recent observational and theoretical studies have indicated that young
SNR are likely sites of efficient magnetic field amplification with a
fluctuating component reaching levels $\gg B_\infty$.\\ High
resolution X-ray observations of young SNR have revealed the presence
of sharp external rims associated with the blastwave, produced by
synchrotron radiation of relativistic electrons
\citep{gott01,hwan02,long03,vink04,ball05}. As discussed in Section~2
below, the smallness of the rim size implies a downstream diffusion
coefficient for the relativistic electrons close to its Bohm value
\citep{bere03,bere04,pari05} i.e. $D_e \simeq r_{\rm L}
c/3$. \citet{bell01} and \citet{luce01} have investigated in some
details the non-linear development of the cosmic ray streaming
instability and obtained an amplified magnetic field well in excess of
$B_\infty$. These authors have suggested that the tangled character of
the magnetic field implies a CR mean free path of the order of its
Larmor radius (the Bohm diffusion regime). Both Bohm diffusion and a
strong magnetic field amplification seem necessary to push the maximum
cosmic-ray energies up to the CR knee at a few PeV or even up to the
CR ankle at a few EeV \citep{ptus03,ptus05}. This amplification
mechanism then appears essential to our understanding of galactic
cosmic-ray phenomenology and theory.

However, previous theoretical work suffers from various
limitations. For instance, the streaming of cosmic-rays amplifies
forward waves at the same time as it damps backward waves, and the
transfer of energy between these two types of waves has not been
modeled by \citet{bell01}. The numerical simulations using coupled
``Particle in Cells" and magnetohydrodynamics codes suffer from
limited wave number dynamics and a crude modeling of the acceleration
mechanism, questioning the value obtained for the non-linear
saturation level of the magnetic field. The argument in favor of a
Bohm diffusion regime is also phenomenological since no MHD theory has
rigorously predicted such regime yet [see \citet{cass02} and
references therein], although \citet{luce01} did observe
isotropisation of the CR distribution on a timescale of a gyroperiod
in their numerical experiments. Finally, the non-resonant instability
mechanism uncovered by \citet{bell04} has not been taken into account
in previous studies. These points have motivated us to investigate in
some details the generation of turbulence in the shock precursor by
cosmic-ray streaming through both resonant and non-resonant
instabilities \citep{pell05}, hereafter refered as paper I. We have
notably found that the three-dimensional spectrum of upstream
turbulence so excited is anisotropic and scales as $S_{3\rm
d}(k_\parallel,k_\perp)\propto k_\parallel^{-1}k_\perp^{-\alpha-1}$
(with $k_\parallel$ along and $k_\perp$ perpendicular to the shock
normal and to the mean magnetic field), and that the downstream
turbulence is further amplified by the jump through the shock.\\ In
the present work, we propose to use the new results of paper I to
investigate the acceleration of cosmic rays in young SNR. In
particular, we explicitly calculate the diffusion coefficients both
upstream and downstream using Monte-Carlo simulations and analytical
quasi-linear theory. We thus provide a connection between the
prediction of turbulence spectra and the observations of electron
synchrotron radiation. We also apply the results to the Fermi cycle of
cosmic rays. Finally we follow in detail the energy budget, accounting
in particular for the energy spent in amplifying the upstream magnetic
field.

\section{Observations}
\subsection{X-ray and radio observations of the outer rims}
Quite recently the X-ray instrument {\it Chandra} has imaged the
continuum emission (4 to 6 keV) in different young supernova remnants
like Cas A \citep{gott01}, Tycho \citep{hwan02}. The X-ray radiation
originates from very thin sheets behind the blast wave.  The same
conclusion have been drawn by XMM-Newton observations of the Kepler
SNr \citep{cass04}. The derived physical width of the emission region
is less than 4$\arcsec$ or 2 10$^{17}$ cm in Cas A, for a shock
velocity $V_{\rm sh} \simeq$ 5000 km/s, less than 4$\arcsec$ or 1.5 \
10$^{17}$ cm in Tycho for a shock velocity $V_{\rm sh} \simeq$ 4600
km/s. The results for three young SNr (Cas A, Tycho and Kepler) and
two older ones (SN1006 and G347.3-0.5) have been presented in
\citet{ball05}. The effective rim size may however be overestimated
due to projection effects \citep{bere03}. \citet{ball05} has also
shown that this radiation is inconsistent with thermal models in a
uniform medium which predict emission everywhere up to the interface,
with only a slight maximum at the blast wave. A second argument
against thermal models is the rather featureless spectrum in the 1-10
keV band which appears to be inconsistent with the X-ray brightness
[see for instance \citet{cass04} in the case of the Kepler SNr]. The
only other possible source of X-ray radiation is suprathermal
particles. Non-thermal bremsstrahlung (by low-energy suprathermal
electrons) could still be compatible with the observed rims for a low
thermal gas temperature of $\sim 1-2 \ 10^6$ K and a high enough
target density \citep{ball05}. Another natural mechanism is the
synchrotron radiation by ultra-relativistic electrons with energies of
few tens of TeV \citep{vink03,vink04}. The radio synchrotron emission
from GeV electrons is expected to have larger extent as the loss
timescale at these energies is much longer. This seems to be indeed
the case for Cas A \citep{gott01} and SN1006 \citep{long03} (for this
source the rim profile has been plotted in radio and in two X-ray
bands showing a systematic sharpening when shifting to the highest
frequencies). The case of Tycho SNr seems to be more complex, here the
two rims (radio and X-rays) do not track each other in brightness well
\citep{hwan02}. We shall discuss in details the physical consequences
of these observations on electrons and cosmic rays in section
\ref{s:xrayobs}.

\subsubsection{Constraints from the shock precursor}
The shock precursor region is of prime importance as it is the place
where most of the turbulence is expected to be
generated. \citet{acht94} noticed that an upper limit on the precursor
size implies an upper limit on the electron mean free path and a lower
limit on the level of turbulence. Depending on the scale over which
the emissivity drops upstream $\ell_{1/2}$, the electron mean free
path is $\lambda_e \le 3 \ (V_{\rm sh}/c) \ \ell_{1/2}/[a
\cos^2(\theta_{Bn})]$ where $\theta_{Bn}$ is the obliquity angle
between the shock normal and the upstream magnetic field, and $a$
accounts for an emissivity drop; it equals $1/\log(2)$ in an
exponential profile or 3 for a smoother profile where the turbulence
is self-generated.

 With the help of high resolution radio observations, \citet{acht94}
concluded to a magnetic field amplification of about one order of
magnitude compared to the standard interstellar values. This
conclusion is however not correct in case of non diffusive transport
where supplementary information about the number of upstream shock
crossings are required \citep{rago01}. If the magnetic field is purely
turbulent upstream, one can replace in the previous expression
cos$^2\theta_{Bn}$ by 1/3. This leads to upper limits for $\lambda_e$
of the ordrer of $10^{-3} {\rm pc} \ (V_{\rm sh}/3000 \ \rm{km/s}) \
(\ell_{1/2}/0.01 pc)$. For instance, Tycho and Kepler have $\ell_{1/2}
\simeq 7 \ 10^{-3}$ pc, $\lambda_e \le 1.1 \ 10^{-3}$pc, and $V_{\rm
sh} \simeq 5400$ km/s, $\ell_{1/2} \simeq 7 \ 10^{-2}$ pc, $\lambda_e
\le 10^{-2}$pc respectively.  The diffusion coefficient for radio
electrons $D \simeq v \lambda_e/3$ is limited to $3 \ 10^{25-26} \
\rm{cm^2/s}$. The Bohm value $D_{\rm Bohm} \simeq 6 \ 10^{22} \
\rm{(E/1GeV)} \ \rm{(B/1\mu G)^{-1}} \ \rm{cm^2/s}$ even if compatible
with these constraints only provides loose constraints.  In X-rays the
emitted radiation must be asymmetrical due to the magnetic field jump
at the shock front. The observed X-ray profile is strongly dependent
on the projection effects and it appears that most (if not all) of the
X-ray photons should come from downstream \citep{bere03}.\\ To
conclude, it is important to note that the synchrotron interpretation
of the radio and X-ray emission is a natural option to explain the
high resolution images from X-rays and radio telescopes and the
non-thermal spectra by Chandra and XMM-Newton.  But this model still
have some difficulties and should be considered with caution. However,
we adopt in the following the synchrotron mechanism as the dominant
radiative process in the sharp outer rims.

\subsection{Consequences on the particle acceleration process}
\label{s:xrayobs}
\subsubsection{Electrons}
The size of the X-ray rims downstream the blast wave have important
constraints on the relativistic particle transport. For electrons, one
can expect that in the loss limited case the size of the rim is set
fixed either by advection or by diffusion. In fact, the electron
energy is expected to be close to the maximum energy given by $t_{\rm
cool}(E_{\rm max}) = t_{\rm acc}(E_{\rm max}) \simeq \kappa/v_{\rm
sh}^2$; in that case the advection scale $\ell_{\rm adv} = t_{\rm
cool} \ V_{\rm sh}$ should be close to the diffusive scale $\ell_{\rm
diff} = \sqrt{\kappa \ t_{\rm cool}}$ \citep{bere03,vink03}. If $E
\simeq E_{\rm max}$, by comparing the advection length with the rim
sizes inferred from the observations, one deduces typical magnetic
field strengths of the order of a few 100 $\rm{\mu G}$ in most of all
young SNR cited above \citep{ball05}, pointing towards an important
magnetic field amplification at the shock precursor by the accelerated
particles themselves. Assuming a diffusion coefficient $D = \alpha \
D_{\rm Bohm} \simeq \alpha \times 6 \ 10^{24} \ \rm{(E/10 \ \rm{TeV})} \
\rm{(B/100\ \rm{\mu G})^{-1}} \ \rm{cm^2/s}$, with $\alpha \ge 1$, the
previous constraints impose $\alpha$ between 1 and 10 \citep{pari05}.

\subsubsection{Cosmic-rays}
In the previous paragraph, we have found that high energy electrons
accelerated close to the Bohm value produce synchrotron X-ray profiles
compatible with the most recent Chandra observations of young
SNr. Likewise, protons and heavier nuclei are prone to diffusive shock
acceleration in this highly disordered magnetic field. However, one
may expect protons energies to be much higher than few tens of TeV and
have Larmor radii allowing them to explore the largest turbulent
scales. At these wavelengths no information is available on the
characteristics of the turbulence yet, i.e. the spectral index, the
turbulence level. Even if the diffusion coefficient is close to its
Bohm value at the energy where the electrons produce synchrotron
photons in the 4-6 keV range, there is no observational evidence that
it remains true at higher energies. In most of the previous approaches
\citep{bell01,ptus03,bere03,bell04} only an heuristic argument was
used, i.e. the particle mean free path is expected to match the Larmor
radius in a completely disordered magnetic field. Numerical
experiments have however shown that in the limit of a strongly
turbulent magnetic field, in the case of a Kolmogorov turbulence, the
diffusion coefficient approaches its Bohm value at large rigidities
only, i.e. for Larmor radius close to the maximum turbulence scale
\citep{cass02}.  \citet{pari05} using the X-ray observations from a
sample of young SNR have investigated all possibilities testing the
spatial transport of cosmic-rays in an isotropic turbulence. They
conclude that spectral turbulence indexes smaller than 3/2 (index of
the 1d spectrum) are to be rejected as they lead to a diffusion
coefficient at the maximum proton energies smaller than the Bohm
diffusion coefficient.  Even if the diffusion regime is close to Bohm,
the authors did not find maximum cosmic-ray energies beyond the CR
knee ($\simeq 3 \ 10^{15}\,$eV), questioning the capability of young
SNr to produce the high energy part of the galactic CR spectrum.

\section{Summary of the results of paper I}
\label{S:3}
In order to discriminate between the different transport regimes
described above, we have investigated in paper I in greater detail the
turbulence spectrum that results from cosmic-ray interactions in the
upstream medium. This work has shown that the turbulence growth occurs
as a result of two instabilities. At large distances from the shock
front, the non-resonant instability \citep{bell04} that results from
the non-zero return current in the thermal plasma, amplifies the
turbulence at short wavelengths (as compared to the typical cosmic-ray
Larmor radius at that distance from the shock front). At closer
distances, cosmic-rays of Larmor radius $r_{\rm L}$ excite resonantly
forward turbulent modes of wavenumber $k_\parallel = 1/r_{\rm L}$
\citep{bell01}. These instabilities govern the growth of the parallel
forward modes and the concomittant damping of the backward parallel
spectrum. While the non-resonant growth saturates when energy is
redistributed through non-linear transfers as fast it is input, the
resonant instability is quenched by advection. The non-resonant
instability that leads to turbulent spectrum $S(k_\parallel)\propto
1/k_\parallel^2$ acts at larger distances from the shock front than
the resonant instability. This latter then supersedes the former,
leading to a turbulent spectrum at the shock front
$S(k_\parallel)\propto 1/k_\parallel$. This one-dimensional spectrum
is related to the three-dimensional spectrum by
$S(k_\parallel)\,\equiv\,(2\pi)^{-2}\int{\rm d}^2k_\perp S_{\rm
3d}(k_\parallel,k_\perp)$, and is normalized according to $\int {\rm
d}k_\parallel S(k_\parallel)\,=\, 2\pi \delta B^2/B_\infty^2$, where
$\delta B$ and $B_\infty$ denote respectively the turbulent component
and the ISM magnetic field.

On a timescale that is shorter than the instability growth time and
the advection time, energy is distributed by resonant three-wave
interactions in the perpendicular direction $k_\perp$. This
redistribution leads to anisotropic three-dimensional turbulent
spectra $k_\perp S_{\rm 3d}\propto
k_\perp^{-\alpha}k_{\parallel}^{-\beta}$, with $\beta=1$ as before,
and $\alpha=(7-2\beta)/3=5/3$ if the energy transfer rate of the
cascade is a constant independent of scale
\citep{gold95,galt00,GPM}. This also implies a relationship between
perpendicular and longitudinal wavenumbers: $k_\parallel L_{\rm
max}\propto (k_\perp L_{\rm max})^{2/3}$, where $L_{\rm max}$ is the
maximum turbulence scale.  These results have been found in agreement
with recent numerical simulations \citep{cho00,maro01}. \\ It has also
been shown in paper I that interactions between two Alfv\'en and one
slow magnetosonic waves provide efficient transfer of energy along the
longitudinal direction, which does not arise when only interactions
between three Alfv\'en waves are considered. This allows to maintain
the backward parallel spectrum close to the level of the amplified
forward spectrum. It has also been pointed out that the non-resonant
instability induces a left-right symmetry breaking which provide
suitable ground for further amplification by a dynamo. This aspect
however has not been studied in detail and is left for further work.\\
The turbulence that is generated by cosmic ray interactions is
compressed along the shock normal at shock crossing. In particular,
the perpendicular magnetic field components are amplified by the shock
compression ratio r while the longitudinal component remains unchanged. 
Correspondingly, the turbulent modes wavenumbers $k_\parallel$ are increased 
by the same factor $r$, which corresponds to a similar decrease of the 
turbulence coherence length in that direction, while perpendicular modes 
are unchanged. The downstream turbulence is thus even more anisotropic. 
In the following, we discuss the implications of these turbulent spectra 
with respect to cosmic ray transport at the shock front, hence with respect to Fermi
acceleration.

\section{Diffusion coefficients}
\label{S:diff}
The diffusion coefficients due to resonant pitch angle scattering of
cosmic rays with turbulent modes can only be derived analytically in
the limit of weak turbulence (quasi-linear theory, Jokipii
1966). Since one expects a high turbulence level in the vicinity of
the shock front, the diffusion coefficients have to be computed
numerically using Monte-Carlo techniques as described in
\citet{cass02}.  The main results of this paper are that the pitch
angle frequency follows the same scaling as the quasi-linear theory
(hereafter QLT) namely $\nu_{\rm s}/\omega_{\rm L} \sim \eta
\rho^{\beta -1}$ for $\rho <1$, where $\beta$ is the index of the
turbulence spectrum, $\nu_{\rm s}=1/\tau_{\rm s}$ is the scattering
frequency (and $\tau_{\rm s}$ the scattering time), and $\omega_{\rm
L}$ the Larmor frequency. For $\rho > 1$ the ratio $\nu_{\rm
s}/\omega_{\rm L}$ decays as $1/\rho$. The reduced rigidity is defined
as:
\begin{equation}
\rho \,\equiv\, {2\pi r_{\rm L}\over L_{\rm max}}\,=\,k_{\rm min}r_{\rm L}\ ,
\end{equation}
with $L_{\rm max}$ the maximum length of the turbulence and $k_{\rm
min}$ the associated minimum wavenumber. Thus, the parallel diffusion
coefficient $D_{\parallel} = (1/3)v^2 \tau_{\rm s} \propto
\epsilon^{2-\beta}$ for small $\rho$ ($\epsilon$ is the particle
energy) and $\propto \epsilon^2$ for large $\rho$, $v\simeq c$ is the
cosmic-ray velocity. The transverse diffusion coefficient $D_{\perp}$
has an unusual scaling which neither corresponds to that from
quasilinear theory or to the Bohm coefficient, as it is rather
controlled by field line chaos, leading to $D_{\perp} =
\eta^{2+\varepsilon}D_{\parallel}$ (with $\varepsilon \sim 0.3$).  In
numerical experiments, there is no observed Bohm scaling, only a Bohm
maximum for the scattering frequency when $\rho \sim 0,1$ and $\eta
\sim 1$.
 
These scaling laws are modified when anisotropic turbulence is
generated, as will be seen in the next section. We have performed new
numerical simulations of particle diffusion in turbulent magnetic
fields, paying particular attention to the anisotropic nature of the
turbulence expected in the vicinity of the shock front. We have
decomposed the three-dimensional turbulence on separate discrete grids
of perpendicular and parallel wavenumbers. This and the required
number of Monte-Carlo runs result in rather long computation times. We
have also derived the diffusion coefficients using the quasi-linear
theory perturbative methods and used these to interpret the numerical
diffusion coefficients, as discussed below.

Assuming the existence of a mean field component $\langle B\rangle$,
one defines the pitch angle diffusion frequency $\nu_s = \tau_{\rm
s}^{-1} = \langle \Delta \alpha^2\rangle/\Delta t$, where $\alpha$ is
the pitch angle with respect to the mean field $\langle \vec B\rangle$
and $\Delta \alpha$ the random jump of $\alpha$ during $\Delta t$,
with $\tau_c \ll \Delta t \ll \tau_s$. The correlation time $\tau_c$
is of order the Larmor time $\tau_L$ and the latter strong
inequalities hold only in weak turbulence regime or in strong
turbulence regime for $\rho \ll 1$.
The function $g\equiv \nu_{\rm s}/\omega_{\rm L}$ can be expressed in
terms of the three-dimensional turbulent spectrum [see \citet{cass02}
and appendix A] as:
\begin{equation}
\label{Eq:g}
g\, =\, \int\frac{{\rm d}^3 k}{(2\pi)^3}\, S_{3\rm
d}(k_{\parallel},k_{\perp})R(\vec k, \vec p) \ ,
\end{equation}
where $\Delta \vec{x}(\tau)$ and $\omega(k)$ stand for the variation
of the unperturbed particle trajectory within a timescale $\tau$ and
for the mode pulsation respectively. $R(\vec k, \vec p)$ is the
resonance function defined by:
\begin{equation}
\label{Eq:R}
R(\vec k, \vec p) \equiv \int_0^{\infty}{\rm d}(\omega_{\rm L} \tau)\,
e^{\left[i \vec{k}.\Delta \vec{x}-i\omega(k).\tau\right]} \cos(\omega_{\rm L} \tau) \ .
\end{equation}
A detailed derivation of the function $g$ including estimates in the
strong turbulence regime can be found in the appendix of
\citet{cass02}. The Landau-synchrotron resonances in Eq.(\ref{Eq:R})
are obtained by inserting the gyro-motion, and we get
\begin{eqnarray} 
\label{Eq:gn}
R(\vec k, \vec p) & = & \sum_{n =
-\infty}^{+\infty}J_n^{2}(k_{\perp}r_{\rm L} \sin\alpha) \times
\nonumber \\ & & \left\{ \delta\left[k_{\parallel}r_L\mu -
(n+1)\right]+\delta\left[k_{\parallel}r_L\mu + (n-1)\right]\right\} \
.
\end{eqnarray}
The expression obtained for linearly polarized Alfv\'en waves in
Eq.(\ref{Eq:gn}) is identical to other derivations
\citep{schl02,yan02}.

\subsection{Upstream diffusion coefficients}
\subsubsection{Modified weak turbulence spectrum}\label{sec:weak}
We start with a model of weak MHD turbulence \citep{galt00} upstream
modified by the streaming instability of cosmic-rays. In that case
(see paper I), the 3D turbulence spectrum reads:
\begin{equation}
\label{Eq:S3Dweak}
S_{3\rm d}(k_{\parallel},k_{\perp}) = S_0 \ u_{\parallel}^{-1} \ u_{\perp}^{-3} \ ,
\end{equation}
where $u_{\parallel} = k_{\parallel} \ \ell_{\parallel}$ and
$u_{\perp} = k_{\perp} \ \ell_{\perp}$. Out of generality we leave the
parallel $\ell_{\parallel}$ and perpendicular $\ell_{\perp}$ magnetic
field maximal lengths unspecified; to make contact with previous
notations, in isotropic turbulence they would correspond to the
maximal length $L_{\rm max}$ introduced earlier. For a given Larmor
radius $r_{\rm L}$ we have thus two rigidities $\rho_{\parallel}$ and
$\rho_{\perp}$.

Since we are mostly interested in the scaling of the diffusion
coefficients with rigidity or Larmor radius, we drop the numerical
prefactors for convenience. Inserting Eq.(\ref{Eq:S3Dweak}) into
Eq.(\ref{Eq:gn}) we find \\
\[
g \,\propto\, \sum_n |n\pm 1| \int_1^{\infty} {\rm d}u_{\perp} \
\left[{J_{n}(u_{\perp} \rho_{\perp})\over u_{\perp}}\right]^2 \ .
\]
Only the harmonic $n = 0$ contributes to the integral
significantly. The function $g$ is then found to be constant for
$r_{\rm L} \ll \ell_{\perp}$, i.e. the pitch-angle scattering time is
proportional to the particle energy. In other words, a turbulence
spectrum $S\propto 1/k_\parallel$ as generated by cosmic-ray
interactions in the shock precursor leads to Bohm diffusion with
$D_\parallel\sim r_{\rm L}c$.

\begin{figure}
\includegraphics[width=0.5\textwidth]{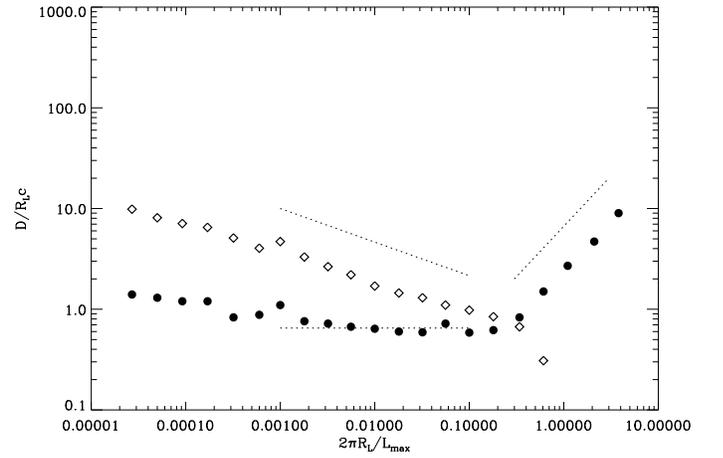}
\caption{Parallel (solid dots) and perpendicular (diamonds) diffusion
coefficients in units of $r_{\rm L}c$ as a function of rigidity $2\pi
r_{\rm L}/L_{\rm max}$, $L_{\rm max}$ denoting the largest scale of
the turbulence. The turbulence level is $\eta=0.95$ and the turbulent
spectrum $k_{\perp}S_{3\rm d}\propto
k_\perp^{-5/3}k_\parallel^{-1}$. The Bohm diffusion regime is clearly
apparent for the parallel diffusion coefficient. The dotted lines have
slope $-1/3$, 0 and 1. On can see a slight offset in the ordinates of
the data points between the two regimes $\rho < 10^{-3}$ and
$\rho>10^{-3}$; this offset is a numerical artifact, see text for
details.}
\label{fig:1}
\end{figure}

This result agrees quite well with the numerical computation of the
diffusion coefficient in the high turbulence level limit, shown in
Fig.~\ref{fig:1} for $\eta=0.95$, where $\eta\equiv \delta B^2/(\delta
B^2 + \langle B \rangle^2)$. As the turbulence level decreases, the
scaling remains the same but the numerical prefactor for the diffusion
coefficients increases, see also \citet{cass02}. This figure reveals a
slight offset in the ordinates of the data points between the two
regimes $\rho<10^{-3}$ and $\rho>10^{-3}$, which is a numerical
artifact. In order to obtain data points down to $\rho\sim10^{-5}$
when only a limited dynamic range in wavenumber is allowed by
computing limitations, we have conducted two separate simulations, one
for each rigidity range, and plotted all results together. Data points
at $\rho<10^{-3}$ have been obtained from a simulation with the full
turbulence spectrum covering five orders of magnitude, i.e. $k_{\rm
max}/k_{\rm min}=10^5$; the simulation for $\rho>10^{-3}$ has a
turbulence spectrum that is cut-off at high wavenumbers but with the
same $k_{\rm min}$ as the previous simulation, and $k_{\rm max}/k_{\rm
min}=10^3$. All rigidities have resonant modes with which to interact
and computational time is saved by removing the high wavenumber modes
(small spatial scales as compared to the Larmor radius) that have a
negligible impact on the particle trajectory. This produces a slight
offset in the diffusion coefficient, which are thus measured to
$\sim10\,$\%; however, the trend of a constant $D_\parallel/r_{\rm
L}c$ is clearly seen in both regimes $\rho<10^{-3}$ and
$\rho>10^{-3}$.

Note also that, in agreement with the above calculation, the scaling
of $D_\parallel$ is insensitive to the slope of $S_{3\rm d}$ in the
perpendicular direction $k_\perp$. On contrary, the perpendicular
coefficient is sensitive to the perpendicular cascade, $D_\perp\propto
r_{\rm{L}}^{1/3}$.

\subsubsection{Modified Goldreich-Sridhar spectrum}
The second turbulence model we shall consider is the Goldreich-Sridhar
model of strong MHD turbulence \citep{gold95} modified by the
streaming instability (see paper I). The three-dimensional turbulence
spectrum then reads:
\begin{equation}
\label{Eq:S3Dstrong}
S_{3\rm d}(k_{\parallel},k_{\perp}) = S_0 \
u_{\parallel}^{-1} \ u_{\perp}^{-8/3} \
f\left(\frac{u_{\parallel}}{u_{\perp}^{2/3}}-1\right) ,
\end{equation}
where $f(x)$ is a smooth function that peaks around $x=0$ and vanishes
at infinity. If we approximate $f$ with a Dirac distribution and
insert Eq.(\ref{Eq:S3Dstrong}) into Eq.(\ref{Eq:gn}), we obtain $g
\propto \omega_{\rm L}^{-1}$ attenuated by $J_n^2(a_n)$ with argument
$a_n \equiv (\rho_{\parallel}/|n\pm1|)^{-3/2} \ \rho_{\perp}$. Since
the resonance condition corresponds to $u_{\parallel} \sim
1/\rho_{\parallel}$, the Golreich-Shridar selection for the transverse
modes leads to $u_{\perp} \sim u_{\parallel}^{3/2} \sim
\rho_{\parallel}^{-3/2} \gg 1$, hence for short transverse
wavelengths. Furthermore, for $\rho_{\parallel} \sim \rho_{\perp} \ll
1$, the argument $a_n\gg1$ and the large argument limit of Bessel
functions gives:
\begin{equation}
\label{Eq:normstrong}
g \sim \eta \rho_{\parallel}^{5/2}/\rho_{\perp} \ .
\end{equation}
We obtain identical results if $f$ is approximated by a step function
or an exponential form.

In this anisotropic turbulence, whose spectrum is constrained by the
relation between $k_\parallel$ and $k_\perp$, the parallel diffusion
coefficient $D_\parallel$ shows a different scaling than in
unconstrained turbulence. On a formal level, this difference is
related to the fact that the relation between $k_\parallel$ and
$k_\perp$ no longer permits to treat perpendicular and parallel
wavenumbers separately in the integrals. The QLT result that we
obtained in Eq.~(\ref{Eq:normstrong}) indicates that the perpendicular
spectrum at short wavelengths influences the pitch angle scattering
frequency.

\begin{figure}
\includegraphics[width=0.5\textwidth]{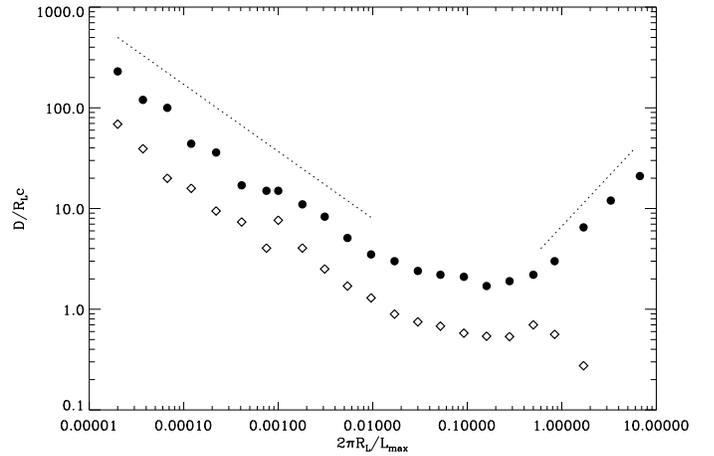}
\caption{Parallel (solid dots) and perpendicular (diamonds) diffusion
coefficients in units of $r_{\rm L}c$ as a function of rigidity $2\pi
r_{\rm L}/L_{\rm max}$. The turbulence level is $\eta=0.95$ and the
turbulent spectrum $k_{\perp}S_{3\rm d}\propto
k_\perp^{-5/3}k_\parallel^{-1}$ with the constraint that
$(k_{\perp}L_{\rm max})^{2/3} > k_\parallel L_{\rm max}$. The dotted
lines have slope $-2/3$ and 1, and indicate that the diffusion
coefficients at low rigidities scale as $r_{\rm L}^{1/3}$ as in
isotropic Kolmogorov turbulence.}
\label{fig:2}
\end{figure}

In the numerical calculation whose results are presented in
Fig.~\ref{fig:2}, we have approximated the function $f(x)$ with a step
function that vanishes outside of the interval $[-1,+1]$ and used, as
before, a turbulence level $\eta=0.95$. We find that the parallel
diffusion coefficient no longer respects the Bohm scaling, but rather
follows the perpendicular coefficient and scales as in Kolmogorov
diffusion, $D_\parallel \propto \rho^{1/3}$. This behavior, which does
not agree with the above quasi-linear estimate
Eq.~(\ref{Eq:normstrong}), may be understood in the following way.

The quasi-linear calculation indicates that diffusion in this
contrainted turbulence cannot be induced by resonant interactions in
the longitudinal direction $k_\parallel r_{\rm L}\sim1$. Indeed, if
this resonance condition is satisfied, the Goldreich-Shridar
prescription $k_{\parallel}l_\parallel\propto
k_\perp^{2/3}l_\perp^{2/3}$ indicates that $k_\perp r_{\rm L}\gg 1$
and this large argument of the Bessel function in Eq.~(\ref{Eq:gn})
then kills the contribution of these interactions in $g$. Since the
parallel and perpendicular diffusion coefficients are found to satisfy
$D\propto r_{\rm L}^{1/3}$, as was observed in unconstrained
turbulence for $D_\perp$, the numerical simulation rather indicates
that diffusion is now controled by perpendicular wavenumbers,
i.e. $k_\perp r_{\rm L}\sim 1$. If indeed those interactions dominate
and the correlation time remains of the order of the Larmor time, the
discussion in Appendix~A shows that the observed behavior can be
reproduced. In this description, pitch angle scattering in the
parallel direction takes place when the particle jumps from one
magnetic field line to another and the orientation of these field
lines differ from one another. Diffusion along the parallel direction
is then controled by perpendicular diffusion, as is apparent in the
results of the numerical simulations shown in Fig.~\ref{fig:2}.


Following the discussion of Appendix~A, we finally obtain:
\begin{equation}
\label{ }
\frac{\nu_s}{\omega_L} \sim \eta \rho_{\perp}^{2/3} \ .
\end{equation}
It is interesting to note that the exponent does not depend on the
turbulence spectral indices $\alpha$ and $\beta$ in a first
approximation (see Appendix~A).

For $\rho_{\parallel} > 1$, the arguments of the Bessel function are
small and we find again the same result as in the isotropic case,
namely a parallel diffusion coefficient increasing like
$\rho_{\parallel}^2$.  Interestingly, the above diffusion coefficients
should be expected to apply to propagation in the interstellar medium
as well if the streaming instability of CR along the mean magnetic
field is the dominant source of turbulence generation. A detailed
investigation of this issue is postponed to a future work.

\subsection{Downstream diffusion coefficients}
Downstream of the shock the parallel and perpendicular coherence
lengths are different: they are related to the coherence length
upstream $L_{\rm max}$ by $\ell_{\perp} = L_{\rm max}$ and
$\ell_{\parallel} = L_{\rm max}/r$. The turbulence transforms in a
non-trivial way, since the magnetic field modes components that are
perpendicular to the shock normal are amplified by $r$, and these
modes may be associated with wavenumbers that are neither completely
parallel nor completely perpendicular. The overall turbulence magnetic
field is nevertheless amplified by $R=\sqrt{(1+2r^2)/3}$ after
statistical averaging, while the coherent magnetic field, which was
assumed parallel, is left unmodified. Hence the turbulence amplitude
$\eta$ changes from upstream to downstream by the above amplification
factor $R$ if $\eta\ll1$, or remains the same if $\eta\sim 1$.\\
The effective value of the compression ratio explored by the particles depends on their 
rigidity in the case of a strongly modified shock. Only the CRs of highest energy should explore the total 
compression ratio, $r_{\mathrm{tot}}$ while low-energy particles explore the much lower 
compression ratio of the subshock, $r_{\mathrm{sub}}$, and the electrons of highest energy explore a region across the 
shock with an intermediate effective compression ratio $r_{\mathrm{el}}$. The fluid quantities (density, velocity, thermal pressure, 
tangential magnetic field) are all subject to a jump dependent on $r_{\mathrm{sub}}$ at the shock precursor. For instance the magnetic
field compression $\mathrm{R}$ ratio should be expressed in terms of $r_{\mathrm{sub}}$ instead of $r$. Typical values are 
$r_{\mathrm{sub}} \simeq 2-3$ and $r_{\mathrm{tot}} \simeq 7-10$ \citep{bere96,byko04}.
However, in section \ref{S:CRacc} we will show that a consistent modelling of young SNr with a strong magnetic field amplification 
by the streaming of CR should include both the CR backreaction on the shock structure and the CR distribution steepening produced 
while generating the turbulence. The last effect will tend to diminish the CR pressure and to reduce the non-linear shock 
modification. One may expect in that case $r_{\mathrm{sub}}$ not to be strongly different from the the test particle case.\\
Mainly for the previous reason and also as they would only complicate the formal treatment with no significant change in 
the results we shall not go into such details in this paper. (Accounting for a different compression ratio $r_{\mathrm{sub}}$ 
and $r_{\mathrm{el}}$ would lead to a variation of the order of 30\% for the deduced value of $E_{\mathrm{p,max}}$ as already 
stressed in \citet{pari05}.) In any case, we found that a situation with explicitly different values of the compression ratio 
for nuclei and electrons in the case of a strongly modified shock results in an intermediate solution between the test-particle 
case (no precursor nor sub-shock, and $r=4$) and the pure $r = 10$ case expected in strongly modified shocks (see \citet{byko04}). 
We shall thus confine the study below to such idealised cases, using only one (``universal'') compression ratio $r$ for all particles 
and the fluid quantities.\\
The transport properties differ in compressed and uncompressed
turbulence, see for instance \citet{lemo05} for the case of
ultra-relativistic shocks. In the following
Figs.~\ref{fig:3},\ref{fig:4} we show the diffusion coefficients of
particles propagating in compressed turbulence (with $r=4$) for the
same cases of turbulence spectra as shown in
Figs.~\ref{fig:1},\ref{fig:2}.

\begin{figure}
\includegraphics[width=0.5\textwidth]{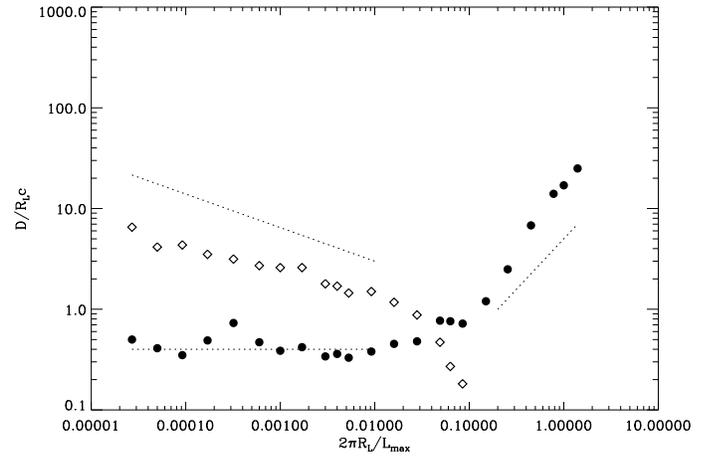}
\caption{Same as Fig.~\ref{fig:1}, i.e. a turbulence spectrum
$k_{\perp}S_{3\rm d}\propto k_\perp^{-5/3}k_\parallel^{-1}$ and
$\eta=0.95$ but accounting for the compression of the turbulence
through the shock.}
\label{fig:3}
\end{figure}

\begin{figure}
\includegraphics[width=0.5\textwidth]{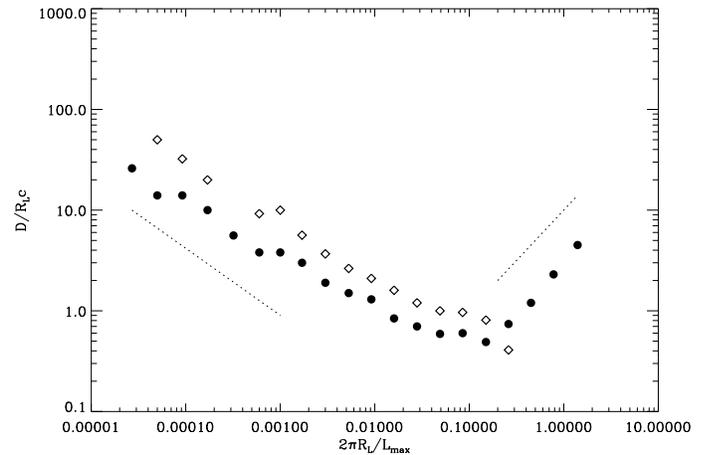}
\caption{Same as Fig.~\ref{fig:2}, i.e. assuming that
$(k_{\perp}L_{\rm max})^{2/3} > k_\parallel L_{\rm max}$, and
accounting for the compression of the turbulence through the shock.}
\label{fig:4}
\end{figure}

The comparison of these figures reveals that the main effect of
compressed turbulence comes through the reduction of the parallel
coherence length: in compressed turbulence, the scaling of
$D_\parallel$ with respect to rigidity $2\pi r_{\rm L}/L_{\rm max}$
remains the same as in uncompressed turbulence, but the curve is
shifted to lower values of the rigidity. Hence if one plotted
$D_\parallel$ versus $r\times2\pi r_{\rm L}/L_{\rm max}$, most of the
discrepancy would disappear. One also observes that the parallel, and
to some a smaller extent the perpendicular diffusion coefficients are
decreased to a lower value. The difference in turbulence level between
downstream and upstream is probably too small to account for this
difference, since $\eta=0.95$ upstream and $\eta=0.98$ downstream.  In
the case of a modified Goldreich-Sridhar spectrum, a reduction factor
of $\simeq 4^{2/3}$ of the maximum turbulence length can explain this
difference.

Another consequence of magnetic field amplification and compression at
the shock front is that the ratio of the downstream Alfv\'en velocity
$V_{\rm Ad}$ to the downstream fluid velocity $V_{\rm d}$ is unlikely
to be much smaller than unity as in the upstream medium far from the
shock front. In detail, the downstream Alfv\'enic Mach number ${\cal
M}_{\rm Ad}\equiv V_{\rm d}/V_{\rm Ad}$ is expressed in terms of the
upstream Alfv\'enic Mach number ${\cal M}_{\rm Au}$ as:
\begin{equation}
{\cal M}_{\rm Ad}\,\simeq\, r^{-3/2}{\cal M}_{\rm Au}\ ,
\end{equation}
due to compression of the magnetic field and density by $\simeq r$.
With $r\simeq 4$ and ${\cal M}_{\rm Au}\sim 10$ [see
Eq.~(\ref{eq:1ma}) below], one finds $V_{\rm Ad}\sim V_{\rm d}$. In
this case, second-order Fermi acceleration should be taken into
account during the Fermi cycle. The stochastic momentum diffusion
coefficient is easily expressed in the relativistic limit as
\begin{equation}
\label{Eq:Dee}
D_{pp} = {1 \over 3} \ p^2 \ \left({V_{\rm A} \over c}\right)^2 \
\nu_{\rm s} \ = {1 \over 9} \ p^2 \ {V_{\rm A}^2 \over D_{\parallel}}
\ ,
\end{equation}
and $p = \epsilon/c$ is the particle momentum. The Fermi second order
acceleration timescale, $t_{\rm FII} = (D_{pp}/p^2)^{-1} \propto
D_{\parallel} $ is then controled by the downstream stochastic
diffusion coefficient. This contribution of stochastic reacceleration
downstream is discussed in Section~\ref{S:Fermi}.

\section{Revised Fermi cycle in shocks with magnetic field amplification}
\label{S:Fermi}
Each relativistic particle streaming upstream of the shock front
generates shear Alfv\'en waves. Accounting for the possibility of
strong turbulence, the energy loss in generating the waves is
\begin{equation}
\label{PL }
P(\epsilon) = -{1\over 3}{v^2 \over c^2} V
\left(\frac{\partial}{\partial x} \log f\right) \epsilon \ ,
\end{equation}
(see Appendix~B), $V$ the velocity of the scattering centers upstream is
close to the local Alfv\'en velocity $V_{\rm{Au}}$.

 This general expression can be specified upstream where the
distribution function is governed by the following transport equation,
assuming a stationary state:
\begin{equation}
\label{eq:dce}
-V_{\rm{u}} \frac{\partial}{\partial x} f + \frac{\partial}{\partial
 x}D_{\rm{u}} \frac{\partial}{\partial x} f=0 \, ,
\end{equation}
The power loss, in most of the upstream region where $f \gg
f_{\rm{u}}$, is then given by
\begin{equation}
\label{ }
P(\epsilon) = -{1\over 3}{v^2 \over c^2}
V\frac{V_{\rm{u}}}{D_{\rm{u}}} \epsilon \, .
\end{equation}
The average duration of a half Fermi cycle upstream is
$4D_{\rm u}/(V_{\rm u} v)$. A particle that undergoes such wandering
and thus contributes to turbulence generation spends an average amount
of energy $-{4 \over 3}{V v\over c^2}$. The energy gain per Fermi
cycle $G \equiv 1 + \xi$ is thus diminished for two reasons, first by
the fact that the particles experience a reduced velocity difference
between scattering centers $V_{\rm{u}}-V_{\rm{d}}-V$ and moreover
loose energy by exciting turbulence; then
\begin{equation}
\label{Eq:xi}
 \xi ={4\over 3} \frac{V_{\rm{u}}-V_{\rm{d}}}{v} -{4\over
 3}\left(\frac{V}{v}+\frac{Vv}{c^2}\right) \ .
\end{equation}

Since the escape probability per cycle is unaffected, $P_{\rm esc} =
4V_{\rm d}/v$, the amplification of upstream turbulence and the
concomittant reduction of the energy gain steepen the accelerated
spectrum. The spectral index is given by \citep{B78}:
\begin{equation}
\label{eq:1index}
s = 1+\frac{\log 1/P_{\rm ret}}{\log G} \simeq 1 + \frac{P_{\rm
esc}}{\xi} \simeq 1+ \frac{3}{r-1- 2{V\over V_{\rm{d}}}} \ ,
\end{equation}
or \beq s \ \simeq \ 1 + {3 \over (r-1)} \times [1 + 2 {r \over
(r-1)}\frac{V_{\rm{Au}}}{V_{\rm sh}}] \ , \eeq or again \beq s \
\simeq \ 2 + {8\over 3} \ \frac{V_{\rm{Au}}}{V_{\rm sh}} \ , \eeq for
a compression ratio r = 4. From Eq.~(\ref{eq:1index}), we can see that
the effect of wave generation upstream on the CR spectrum is important
once the amplitude of the fluctuations $\delta B$ becomes much larger
than the ISM magnetic field. Let us quantify this remark. The index
can be expressed in terms of the inverse of the {\it turbulent}
(upstream) Alfv\'en Mach number \beq
\label{eq:1ma}
M_{\rm A} = {V_{\rm sh}\over V_{\rm{au}}}\, \simeq 14.3 \
\left(\frac{B}{100 \mu G}\right)^{-1} \ \left(\frac{n}{1 \
\rm{cm}^{-3}}\right)^{1/2} \ \frac{V_{\rm sh}}{10^{-2} c} \ .  \eeq
Instead of an index of 2 for strong shocks, we get for a SNr shock
$V_{\rm{sh}} = 5000 \ \rm{km/s}$, with an amplified upstream magnetic
field of $\simeq 100 \mu$G \citep{bere04,pari05} in a hot ISM medium
$n_{\rm HIM} \simeq 4 \ 10^{-3} \ \rm{cm}^{-3}$, $s \ =\ 2 + 1.5$ and
in a warm ISM medium $n_{\rm WIM} \simeq 10^{-1} \ \rm{cm}^{-3}$, $s \
= \ 2 + 0.35$ respectively. The index appears to be strongly modified
especially in the hot phase.\\

The compression ratio can also be changed with a correction of order
$1/{\cal M}_{\rm{Au}}$ by the modification of the shock by both
CR-pressure gradient and wave-pressure gradient. Note that the thermal
plasma pressure also increases in the precursor because the magnetic
energy cascade towards dissipation scale where cyclotron absorption
takes place. These effects tend to diminish the compression ratio. 
Using Eq. (\ref{eq:1ma}) for $r = 10$ the spectral index $s \simeq 1.65$ 
in the warm ISM is found to be slightly harder than the solution obtained in modified 
shock simulations ($ s \simeq 1.5$, \citet{Ell04}). All these effects are still to 
be integrated in numerical simulations of CR modified shocks including kinetics and 
the effect of a strong Alfv\'en production upstream [see \citet{jone94} for a discussion in
the three fluid approximation].

As already mentioned, the generation of a highly amplified and
compressed magnetic field enhances the Alfv\'en velocity compared to
far upstream in such a way that the Fermi second ordrer process may
become important. However, if only forward waves are generated, the
Fermi second order process does not work. Assuming that the energy is
distributed evenly between forward and backward waves, one can
estimate the mean energy gain through second order acceleration using
Eq.~(\ref{Eq:Dee}):
\beq
\label{Eq:f2}
g_{\rm FII} \simeq \frac{4}{3} \frac{v V_{\rm d}}{c^2} \,
\left[\left(\frac{V_{\rm{Ad}}}{V_{\rm{d}}}\right)^2 \ \frac{1}{p^3 \
\nu_{\rm{sd}}} \frac{\partial \left(p^4 \nu_{\rm{sd}}\right)}{\partial
p} \right]\ .  \eeq The overall particle energy gain/loss during one
cycle transforms to: \beq
\label{Eq:gcyc}
G_{\rm{tot}} = 1 + {4\over 3} \frac{V_{\rm{u}}-V_{\rm{d}}}{v} -
{4\over 3}\left(\frac{V_{\rm{Au}}}{v}+\frac{V_{\rm{Au}}v}{c^2}\right)
+ g_{\rm FII} \ .  \eeq If we note $g_{\rm FI} = 4/3 \
(V_{\rm{u}}-V_{\rm{d}})/v$ the mean energy gain corresponding to the
first Fermi process, then the use of Eq.~(\ref{Eq:f2}) gives: 
\beq
\label{Eq:rg}
{g_{\rm FII} \over g_{\rm FI}} \simeq {3-\kappa \over r-1}
\left({V_{\rm{ad}} \over V_{\rm{d}}}\right)^2 \ , \eeq where $\kappa$
is defined by the energy (rigidity) dependence of the diffusion
coefficient: $D_\parallel \propto D_{\rm Bohm} \rho^\kappa\propto
\rho^{\kappa+1}$, with $D_{\rm Bohm}\equiv r_{\rm L}c$. If the
anisotropy of the turbulence is not constrained by a relation of the
form $k_\parallel \propto k_\perp^{2/3}$, as discussed in
Section~\ref{sec:weak}, then $\kappa=1-\beta$, with $\beta$ the index
of the turbulence power spectrum as a function of $k_\parallel$. In
particular, $\beta=1$ implies $\kappa=0$, hence a Bohm diffusion
coefficient.

 Usually, magnetic field strength up to $500\,\mu$G are derived in
young SNr producing non-thermal X-ray radiation. The previous equation
implies that the second order process can become comparable or even
dominate the regular Fermi process for ISM densities lower than a few
tens (corresponding to the hot ISM phase) for such high magnetic field
strengths. The stochastic process has then to be incorporated in a
self-consistent way in the Fermi cycle in such cases. However, in
order to do so one has to take into account the likely anisotropy of
the turbulence between backward and forward waves and evaluate its
effect on the efficiency of the second order Fermi process. This
investigation remains beyond the scope of this paper and is left for
future work.

\section{Discussion: Cosmic-ray acceleration in SNr}
\label{S:CRacc}
In this section, we aim at deriving the maximum CR energies
produced by the diffusive shock acceleration processes in an 
anisotropic turbulence. We limit our calculations to the young 
SNr showing thin X-ray filaments as observed by the satellites Chandra and 
XMM-Newton (see \citet{ball05}). We postpone to a future work the 
derivation of the overall source CR spectrum produced by SNr during
their whole evolution phases as discussed recently in a similar context
by \citet{ptus05}.\\
In section 4, we have derived the general form of the diffusion
coefficient up- and down-stream. Expressed in terms of the Bohm
diffusion coefficient $D_{\rm Bohm}\equiv r_{\rm L}c$, we get: \bea
\rm{D(\rho)} & = & k\, \rm{D_{Bohm}} \
\left(\frac{\rho}{\rho_{p}}\right)^{\alpha} \ , \rm{for} \ \rm{\rho
\le \rho_{p}} \nonumber \\ & = & k\, \rm{D_{Bohm}} \
\left(\frac{\rho}{\rho_{p}}\right)^{2} \ , \rm{for} \ \rm{\rho >
\rho_{p}} \ , \eea 

From the simulations we infer approximatively $\rho_{\rm{pu}} \simeq
0.2$ upstream and $\rho_{\rm{pd}} \simeq 0.2/r$ downstream, $r$ being
the shock compression ratio. The normalisation factor k is also
different up- and downstream.  In the case of a Bohm type turbulence
upstream, compressed by a factor r downstream (assumed to be 4
hereafter) we have $\rm{k_u} \simeq 2$ and $\rm{k_d} \simeq 4$
respectively (see figures 1 and 3). In case of a Kolmogorov type
turbulence upstream, compressed by a factor r downstream, $\rm{k_u}
\simeq 6$ and $\rm{k_d} \simeq 6/4^{2/3} \sim 2.4$ (see figures 2 and
4). \\ In order to discuss the particle acceleration process in young
SNr quantitatively, all the lengths are compared to the size of the
SNr and all the timescales to the age of the SNr. The maximum scale of
the turbulence up-stream and the particle gyroradius can be expressed
in terms of the SNr shock radius units, namely $\rm{L_{max}} =
\rm{\overline{L}_{max}} \times \rm{R_{sh}}$ $\rm{r_L} =
\rm{\overline{r}_L(E,B)} \times \rm{R_{sh}}$ with,
$\rm{\overline{r}_L} = \rm{\overline{r}_0} \ \rm{E_{PeV}} \
\rm{B}_{100}^{-1}$, $\rm{E_{PeV}}$ is the CR energy in PeV and
$\rm{B}_{100}$ is the magnetic field intensity in 100 $\mu$Gauss
units.  Hereafter, we will restrict our analysis to young SNr in a
free expansion phase for which the forward shock velocity is the
largest.  This leads to a direct relationship between the SNr radius
and age $\rm{R_{sh}} \simeq \rm{V_{sh}} \times \rm{t_{SNR}}$: \beq
\rm{R_{sh}} \simeq \frac{3}{2} \ \rm{pc} \ \frac{\rm{V_{sh}}}{5000 \
  \rm{km/s}} \times \frac{\rm{t_{SNR}}}{300 \rm{yr}} \ , \eeq hence
$\rm{\overline{r}_0} \simeq 2 \ 10^{-2}$ for these values. We now
derive the explicit values of the maximum cosmic-ray energies.

 
Note that downstream, the turbulence may also relax towards ISM values
over a scale $\rm{\ell_{r}}(E)$ \citep{pohl05} producing a diffusion
coefficient $D_{\rm d} \simeq D_{\rm d}(\rm{\ell_r} \rightarrow
\infty) \times \exp(x(\alpha+1)/2\rm{\ell_{r}})$, with $D_{\rm
d}(\rm{\ell_r} \rightarrow \infty)$ the diffusion coefficient in a
uniform medium as before. Relaxation scales $\rm{\ell_r} \ll R_{sh}$
have for average effect to increase the particle residence downstream
and then the acceleration timescale. It is expected that acceleration
through a relaxed turbulence is less efficient in producing high
energy CR. A detailed investigation of this process deserves to be
pushed further.

\noindent \underline{Maximum CR energy in young SNr}:\\
If the relaxation scale is $\rm{\ell_r} \sim \rm{R_{sh}}$ then the
X-ray filaments observed in some young SNr are limited by the
radiative (synchrotron) losses. Several authors
\citep{bere03,bere04,pari05} have derived in a self-consistent manner
the downstream magnetic field by comparing the observed filament size
$\Delta R_{\rm{rim}}$ with combined advection and diffusive lengths
explored by the relativistic electrons during their synchrotron loss
timescale. The magnetic field strengths obtained are of the order of
$300-400\,\mu$G in the youngest SNr like Kepler and Tycho and
$100\,\mu$G in older SNr like SN1006. These estimates however have
assumed an isotropic turbulence up- and downstream.

As discussed in the introduction, the anisotropy considered here has
two main effects. Firstly, if the Goldreich-Sridhar scaling does apply
then diffusion has a Kolmogorov scaling even if the streaming
instability tends to produce a spectrum $\propto
k_\parallel^{-1}$. Secondly, this turbulence spectrum is transmitted
downstream and the scales parallel to the shock normal (in a
quasi-parallel shock configuration) are compressed by a factor close
to $r$. We shall account for both these effects in the following
estimates. 

We first derive the magnetic field downstream using the size of the
X-ray filaments [see \citet{bere04} and \citet{pari05}].  The electron
particle distribution $f(x)$ verifies $V_{\rm{d}}(\partial f/\partial
x)=D\partial^2f/\partial x^2-f/\tau_{\rm{syn}}$.  In this equation,
catastrophic losses for TeV electrons have been assumed. The solution
is of the form $f(x) \propto \exp(-x/\Delta R_{\rm{rim}})$. The
magnetic field downstream then follows from: \beq
\label{Eq:Bd}
\left(\frac{D}{\Delta R_{\rm{rim}}}\right)^2 +
\frac{V_{\rm{d}}}{\Delta R_{\rm{rim}}} = \frac{1}{\tau_{\rm{syn}}} \ .
\eeq The synchrotron loss time $\tau_{\rm{syn}} \simeq (95 \rm{yr}) \
B_{\rm{d100}}^{-3/2} \ E_{\rm{obs-keV}}^{-1/2}$, where
$E_{\rm{obs-keV}}$ is the photon energy in keV units at which the
X-ray filaments are observed and $B_{\rm d100}$ is the downstream
magnetic field in units of 100 $\mu$G. Once the dependence of the
diffusion coefficient on particle energy is known, Eq.~(\ref{Eq:Bd})
leads to a one to one relationship between the downstream magnetic
field and $\Delta R_{\rm{rim}} = \Delta \overline{R}_{\rm{rim}} \
R_{\rm{sh}}$.  The ratio of the upstream to downstream magnetic field
is approximately $1.2 / r$ \citep{pari05}. The maximum cosmic-ray
energy $E_{\rm{pmax}}$ is then calculated by balancing the diffusive
length upstream $D_{\rm{u}}/V_{\rm sh}$ with the shock radius [see
\citet{hill05} for a discussion]. In order to fix the maximum scale of
the turbulence, let us first assume $L_{\rm max} = r_{\rm{L}}
(E_{\rm{pmax}})$. \\
\vspace{0.25cm}

\noindent \underline{Bohm regime}: In the case where diffusion
proceeds according to a Bohm regime upstream, $\kappa = 0$. The
magnetic field downstream is then derived as $B_{\rm d} \simeq
400\,\mu$G for a typical filament size $\Delta \overline{R}_{\rm{rim}}
= 10^{-2}$ and $r = 4$. The results are similar for r=10 in case of a
strongly modified shock.  Such high magnetic fields downstream require
an efficient amplification in the shock precursor (see paper I). Using
the above argument on upstream CR escape, we find maximum cosmic ray
energies $E_{\rm{p max}} \simeq 1.5\,$PeV for $r=4$ or $0.6\,$PeV for
$r = 10$.
\vspace{0.25cm}

\noindent \underline{Kolmogorov regime}: if the upstream turbulence is
of the Goldreih-Shridar type, the diffusion is similar to Kolmogorov,
$\kappa = -2/3$ and Eq.(\ref{Eq:Bd}) leads to a magnetic field of the
order of $B_{\rm d} \simeq 550\,\mu$G for filament size $\Delta
\overline{R}_{\rm{rim}} = 10^{-2}$ and $r = 4$, $B_{\rm d} \simeq 400
\ \mu$G for $r = 10$. Using the above argument on upstream CR escape,
we find maximum cosmic ray energies $E_{\rm{pmax}} \simeq 7\,$PeV
($r=4$) or $\simeq2\,$PeV ($r=10$). 
\vspace{0.25cm}

These estimates suggest that it is possible to accelerate CR up to the
CR knee with magnetic field amplification in the shock precursor for
both Bohm and Kolmogorov regimes. This confirms earlier results found
for isotropic turbulence \citep{pari05} even if the typical maximum
energies for CR tend to be slightly higher in case of anisotropic
compressed turbulence. Another difference with the isotropic
turbulence case, is that the Kolmogorov regime can not be ruled-out
here; the rejection condition $D \le D_{\rm{Bohm}}$ is not true
anymore in anisotropic turbulence. However, it seems still challenging
at this SN evolution stage to accelerate the particles up to the CR
ankle [see \citet{ptus05}, \citet{byko01} and \citet{pari04} for
alternative scenario].

The above maximum CR energies are probably slightly overestimated
since the diffusion coefficient $D\propto \rho^2$ at high rigidities
and the hypothesis $L_{\rm{max}} = r_{\rm{L}} (E_{\rm{pmax}})$ is
optimistic. In order to estimate this uncertainty, we use a
supplementary relation to fix $L_{\rm{max}}$. For instance, the
condition $\tau_{\rm{acc}}(E_{\rm{emax}}) =
\tau_{\rm{sync}}(E_{\rm{emax}})$ at the maximum electron energy
provides a one to one relationship between $L_{\rm{max}}$ and the
magnetic field.  For the youngest SNr, the typical synchrotron cut-off
frequency is of the order of $1\,$keV. Downstream magnetic fields in
both regimes are still of the order of $400-500 \ \mu$G.  In the
Kolmogorov regime, $\overline{L}_{\rm{max}} \simeq 10^{-2}$ and the
maximum CR energies cannot lie well beyond $0.3\,$PeV as a result of
the scaling $D\propto \rho^2$ at high rigidities. 

To summarize, the above estimates give as an order of magnitude for the
maximum energy of cosmic rays: $E\sim Z\times (0.3-3)\,$PeV in young
SNr showing X-ray filaments if the magnetic field is amplified by the 
streaming instability.

This paper only considered the free expansion stage of SNr evolution and did not
investigate the overall spectrum, expected to be steep beyond the CR knee \citep{ptus05}. The maximum
CRs energies calculated here are estimates expected once a strong magnetic field
amplification is at work at the shock precusor as deduced from X-ray observations from young SNr.
Further work should take into account the different SNr evolution phases either, as well as the 
different type of SN (core-collapsed or type Ia). A consistent spectrum derivation should account for 
both upstream wave generation and shock smoothing by the cosmic-ray pressure as well as the second order 
Fermi acceleration process downstream. Consequently, a detailed reconstruction of the CR spectrum produced by 
an ensemble of SNr during their evolution is beyond the scope of the paper and will be adressed in future work.

\section{Conclusion}

In this paper II, we have shown that the diffusion coefficients of
cosmic rays can be significantly modified in the precursor of a shock,
especially because of the formation of an anisotropic turbulence
spectrum. 
According to paper I, the streaming instability shapes the power
spectrum of the turbulence according to the scaling $S_{\rm 3d}\propto
k_{\parallel}^{-1}k_\perp^{-\alpha-1}$, while the transverse index
$\alpha$ depends on the modeling of the Alfv\'en cascade. In the
absence of any correlation between parallel and transverse wave
numbers, the spectrum factorizes and the diffusion coefficients are
insensitive to the transverse distribution. The parallel diffusion
coefficient respects a Bohm scaling, $D\propto r_{\rm L}c$. However,
if a correlation {\it \`a la} Goldreich-Shridar arises,
i.e. $k_{\parallel} \ell_{\parallel} \sim (k_{\perp}
\ell_{\perp})^{2/3}$, the results differ from the quasi-linear
prediction. Numerical simulations indicate that the parallel diffusion
coefficient scales with rigidity as in isotropic Kolmogorov
turbulence.  This behavior is understood if pitch angle scattering is
now controled by transverse transport from one field line to another.
In any case these results allow to explain the recent X-ray
observations young Supernova remnants. The diffusion coefficient for
the electrons, in particular, agrees well with the expected size of
the filaments. 

The energy loss per particle that accompanies the generation of
turbulence is calculated. The cosmic-ray energy spectrum is
consequently steepened, which supports the suggestion that the index
at the source is closer to $2.3$ (index obtained in the warm ISM
phase) rather than $2$. A further steepening due to the diffusive
propagation in the interstellar medium would likely leads to an index
close to $2.7$ as long as the diffusion develops with a Kolmogorov
spectrum. Finally the amplification of the magnetic field allows to
push the maximal energy cut-off into the ``knee" region (PeV
energies). Consistent future investigations with the 
cosmic ray data should require the inclusion of the magnetic field amplification 
by the streaming instability, the shock modification by the cosmic ray 
pressure and possibly the effect of stochastic Fermi acceleration downstream 
the shock.

\section{Acknowledgment}
The authors thank an anonymous referee for helpful comments. A.M. acknowledges fruitful discussions 
with J. Ballet, F. Casse and E. Parizot.\\

\appendix
\section{Pitch angle diffusion}

The variations of the pitch angle are governed by a simple stochastic
equation that stems from the projection of the Lorentz equation along
the mean field by taking into account the energy conservation (and
thus the conservation of $p$ and $v$). This gives:
\begin{equation}
\label{SPA}
\dot \alpha = f(t) \equiv \omega_L\left[\cos \phi(t) b_2(t) - \sin
\phi(t) b_1(t)\right] \ ,
\end{equation}
where $\omega_L \equiv Ze\overline B/m\gamma c$, $\phi(t)$ is the
gyro-phase, i.e. $\dot \phi(t) = \omega_L + {\cal O}(b)$ and $\vec b
\equiv \delta \vec B/\overline B$ is the irregularity of the field
experienced by the particle along its trajectory. For a stationary
process, the pitch angle frequency is thus
\begin{equation}
\label{ }
\nu_s \equiv \frac{\langle\Delta \alpha^2\rangle}{\Delta t} =
2\int_0^{\infty} \langle f(\tau)f(0)\rangle\, {\rm d}\tau \ ,
\end{equation}
which can be rewritten as follows:
\begin{equation}
\label{NUS}
\nu_s = \omega_L^2 \int_0^{\infty} \langle\vec b(\tau)\cdot \vec b(0)
\cos \Delta \phi(\tau)\rangle\,{\rm d}\tau \ .
\end{equation}
If the level of turbulence is sufficiently weak, one approximates
$\Delta \phi(\tau) \simeq \omega_L \tau$ and $\vec b(\tau)$ is
expressed in term of the unperturbed trajectory. This leads to the
expression:
\begin{equation}
\label{ }
\nu_s = \omega_L \int \frac{{\rm d}^3k}{(2\pi)^3}
S_{\rm 3d}(\vec k) R(\vec k, \vec p)\ ,
\end{equation}
where $S_{\rm 3d}(\vec k)$ is the 3D correlation spectrum of the field
irregularities normalized such that its integral equals to the
irregularity degree $\eta$:
\begin{equation}
\label{ETA}
\eta \equiv \frac{\langle\delta \vec B^2\rangle}{\langle\vec
B^2\rangle} = \int \frac{{\rm d}^3 k}{(2\pi)^3} S_{\rm 3d}(\vec k) \ ,
\end{equation}
and the resonance function $R(\vec k, \vec p)$ describes the resonant
interaction between the particles and the modes:
\begin{equation}
\label{ }
R(\vec k, \vec p) \equiv \omega_L \int_0^{\infty} e^{i\vec k. \Delta
\vec x(\tau)-i\omega(k)\tau} \cos(\omega_L \tau) \, {\rm d}\tau \ ,
\end{equation}
where $\Delta \vec x(\tau)$ is the variation of the unperturbed
trajectory during a time lapse $\tau$ and $\omega(k)$ the mode
pulsation. This exhibits the Landau-synchrotron resonances of the form:
\begin{equation}
\label{RLS}
R(\vec k, \vec p) \propto \delta\left[\omega(k)
  -k_{\parallel}v_{\parallel} \pm n\omega_{\rm L}\right] \ .
\end{equation}

For an isotropic power law spectrum $S(k) \propto \eta k^{-\beta}$,
\begin{equation}
\label{ }
\nu_s \sim \eta \omega_{\rm L}\rho^{\beta-1}
\end{equation}
where $\rho \equiv 2\pi r_{\rm L}/L_{\rm max}$, $L_{\rm max}$ being
the coherence length of the field. The parameter $\rho$ is the reduced
rigidity and this law holds for:
\begin{equation}
\label{ }
\frac{k_{\rm min}}{k_{\rm max}} \la \rho \la 1 \ .
\end{equation}
Finally the scattering time $\tau_{\rm s} \sim \nu_{\rm s}^{-1}$.

In the case of strong turbulence, the gyro-resonances broaden; it
turns out that the scaling law in terms of the rigidity $\rho$ and of
the irregularity level $\eta$ can be extrapolated (\cite{cass02}) when
the spectrum is {\it isotropic}. In this paper, we show that the
results can be different when the spectrum is {\it anisotropic}. When
there is no correlation between the parallel and the perpendicular
parts of the spectrum (i.e. $S_{\rm 3d} (\vec k) \propto
k_{\parallel}^{-\beta} k_{\perp}^{-q}$ with $q\equiv\alpha+1 > 2$) the
result derived from the quasi-linear theory can be extrapolated to
strong turbulence, giving a diffusion law similar to the isotropic
case. In particular a Bohm scaling is found for $\beta = 1$ (and only
in that case).

The result is quite different when there is a correlation between
parallel and transverse wave numbers as prescribed by Goldreich and
Shridar, i.e. $k_{\perp}\ell_{\perp} \sim
(k_{\parallel}\ell_{\parallel})^{3/2}$. Now the resonance
$k_{\parallel}r_{\rm L} \sim 1$ becomes inefficient because the requirement
to have simultaneously much larger $k_{\perp}$ makes the particle
insensitive to these modes (i.e.  the angular scattering frequency
vanishes). The numerical simulation actually indicates that the
particles interact at $k_{\perp} r_{\rm L} \sim 1$, as shown in
Fig.~\ref{fig:2}. This behavior is not described by the quasi-linear
theory that would give a non vanishing result only if one could obtain
simultaneously $k_{\parallel} r_{\rm L} \sim 1$.  

However, if one assumes that the particle interacts with a spectrum
band such that $k_{\perp} r_{\rm L} \sim 1$ and the correlation time
is a few Larmor times, as is usually the case, Eq.~(\ref{NUS}) gives a
relation of the form:

\begin{equation}
\label{ }
\nu_s \sim \omega_{\rm L}^2 t_{\rm L} \left[k_{\perp}^2 \int {{\rm
d}k_\parallel\over (2\pi)}S_{\rm 3d}(k_{\perp},
k_\parallel)\right]_{k_{\perp}r_{\rm L} = 1} \ ,
\end{equation}
which leads to a result in agreement with the numerical simulation, if
one recalls the relation $3\alpha+2\beta=7$ [see paper~I and
\citet{GPM}]:
\begin{equation}
\label{ }
\frac{\nu_s}{\omega_L} \sim \eta \rho_{\perp}^{2/3} \ .
\end{equation}
 This effect can be associated with a perpendicular diffusion process
associated to encounters with oblique magnetic field lines
encounterings. The particle angular diffusion is then controled by the
perpendicular transport.

\section{Power loss rate per particle}

Let $\gamma_{\rm r}(\vec k)$ be the growth rate of the resonant
instability (see paper I) and $P(\epsilon)$ the power lost by each
particle of energy $\epsilon$ to trigger the turbulence spectrum
$S_{\rm 3d}(\vec k)$. Then, these physical quantities are linked the
following integrals, expressing that the power generation of waves is
equal to the power loss suffered by the particles:
\begin{equation}
\label{A1}
\int P(\epsilon) f(p) \,4\pi p^2{\rm d}p\, =\, -2 \int \gamma_{\rm r}(\vec k)
S_{\rm 3d}(\vec k) \,\frac{{\rm d}^3k}{(2\pi)^3}\, \bar B^2 \ .
\end{equation}
The growth rate is proportional to the gradient of the distribution
function and depends on the pitch angle frequency $\nu_{\rm s}(p, \mu)$. As
seen in section \ref{S:diff}, the spectrum factorizes (except in the
Golreich-Shridar case) so that the scattering frequency is independent
of the transverse spectrum and is given by
\begin{equation}
\label{A2}
\nu_{\rm s}(p, \mu) = \left .\omega_{\rm L}(p) k_{\parallel}
S_{\parallel}(k_{\parallel})\right\vert_{k_{\parallel}=k_{\rm r}(p, \mu)} \ ,
\end{equation}
where $k_{\rm r}(p, \mu) = (\vert \mu \vert r_{\rm L})^{-1}$.  Then
the growth rate is given by the following integral:
\begin{eqnarray}
\label{A3}
\gamma_{\rm r}(k_{\parallel})&\,=\,& \frac{\pi V}{8 \overline B^2} \int
p^2{\rm d}p \int {\rm d}\mu \frac{\omega_{\rm L} }{\nu_{\rm s}}
\left(1-\mu^2\right)\nonumber\\ & &\times \frac{v^2}{c^2} \epsilon
\delta\left(k_{\parallel}r_{\rm L}\mu -1\right) \frac{\partial f}{\partial
x} \ .
\end{eqnarray} 
Inserting eq.(\ref{A3}) and eq.(\ref{A2}) into eq.(\ref{A1}), one obtains

\begin{eqnarray}
\label{A4}
\int P(\epsilon) f(p) 4\pi p^2{\rm d}p &\,=\,& -{V \over 4}\int_{-1}^{+1}
(1-\mu^2){\rm d}\mu\nonumber\\ & &\times \int \frac{v^2}{c^2} \epsilon
\frac{\partial f}{\partial x} 4\pi p^2 {\rm d}p \ .
\end{eqnarray}
This result is in agreement with (\cite{mack82}) who obtained a power
density for Alfv\'en wave generation equal to $V\partial P_{\rm cr}/
\partial x$. One finally obtains the power loss per particle:
\begin{equation}
\label{A5}
P(\epsilon) = -\frac{1}{3} V \frac{v^2}{c^2} \left(\frac{\partial \log 
f}{\partial x}\right) \epsilon \ .
\end{equation}

However a model of CR acceleration at SNr shock to be consistent 
with the CR data should include at once the amplification of turbulence 
upstream and the back reaction of cosmic ray on the shock structure.

\end{document}